\documentclass[12pt]{iopart}

\usepackage{iopams} 
\newcommand{\text}{\mathrm}
\usepackage{color} 
\usepackage{tabularx} 
\usepackage{epsfig}
\usepackage{amssymb} 
\usepackage{graphicx}
\usepackage{wasysym}
\usepackage{psfrag}

\definecolor{bls}{named}{Blue} 
\definecolor{eme}{named}{ForestGreen}

\newcommand{\prima}{^\prime}
\newcommand{\ts}[1]{{\mbox{\scriptsize{#1}}}}
\newcommand{\av}[1]{\left[ #1 \right]_\ts{av}}
\newcommand{\aT}[1]{\left\langle #1 \right\rangle_\ts{T}} 

\makeatletter
\def\moverlay{\mathpalette\mov@rlay}
\def\mov@rlay#1#2{\leavevmode\vtop{%
\baselineskip\z@skip \lineskiplimit-\maxdimen
\ialign{\hfil$#1##$\hfil\cr#2\crcr}}}
\makeatother

\newcommand{\tetra}{\moverlay{\triangle\cr \cdot}}

\begin{document}

\title[Tricolored Lattice Gauge Theory with Randomness]{
{\color{red}T}{\color{eme}r}{\color{blue}i}{\color{red}c}{\color{eme}o}{\color{blue}l}{\color{red}o}{\color{eme}r}{\color{blue}e}{\color{red}d}
Lattice Gauge Theory with Randomness:\\ 
Fault-Tolerance in Topological Color Codes}
\author{Ruben S.~Andrist}
\address{Institute for Theoretical Physics, ETH Zurich, CH-8093 Zurich, Switzerland}

\author{Helmut G.~Katzgraber} 
\address{Department of Physics and Astronomy, Texas A\&M University,
College Station, Texas 77843-4242, USA}
\address {Institute for Theoretical Physics, ETH Zurich, CH-8093 Zurich, Switzerland}

\author{H.~Bombin}
\address{Perimeter Institute for Theoretical Physics, Waterloo,
Ontario N2L 2Y5, Canada}

\author{M.~A.~Martin-Delgado}
\address{Departamento de F{\'i}sica Te{\'o}rica I, Universidad 
Complutense, 28040 Madrid, Spain}

\date{\today}

\begin{abstract}
We compute the error threshold of color codes---a class of topological
quantum codes that allow a direct implementation of quantum Clifford
gates---when both qubit and measurement errors are present. By
mapping the problem onto a statistical-mechanical three-dimensional
disordered Ising lattice gauge theory, we estimate via large-scale
Monte Carlo simulations that color codes are stable against 4.8(2)\%
errors. Furthermore, by evaluating the skewness of the Wilson
loop distributions, we introduce a very sensitive probe to locate
first-order phase transitions in lattice gauge theories.
\end{abstract}



\maketitle

\section{Introduction}
\label{sec:intro}

The study of quantum error correction in topological stabilizer
codes \cite{kitaev:03} burgeoned a magnificent synergy between
quantum computation and statistical-mechanical systems with disorder
\cite{dennis:02-ea}. Quantum error-correction \cite{shor:95,steane:96}
is a method to preserve quantum information from decoherence
by actively detecting and counteracting errors using redundancy.
In topological codes, these error-detecting operations are local and
one can relate the stability of a topological quantum memory to an
ordered phase in a classical statistical model \cite{dennis:02-ea}. The
error threshold---a figure of merit that describes the stability of
a system against local errors and up to which error correction is
feasible---can be identified with the critical point at which the
ordered phase of the underlying statistical mechanical system is lost
due to the influence of quenched disorder.

In most studies, it is assumed that the error-correction procedure
is free of errors. However, error-correction is achieved by
means of applying a set of quantum gates and this procedure can
be flawed as well, leading to the concept of fault-tolerance
\cite{shor:96,knill:96-ea,aharonov:97}.  In a topological quantum
memory, the information is encoded in global degrees of freedom
\cite{dennis:02-ea} and preserved by repeatedly performing local
measurements to keep track of and correct for errors.  To understand
the error resilience, it is convenient to adopt a phenomenological
description of errors \cite{dennis:02-ea}. Errors are assumed to be
nonsystematic and uncorrelated both in space and in time. Therefore,
the error modeling process is parametrized in terms of two error
rates: the qubit error rate $p$ and the measurement error rate
$q$. At each complete step in the syndrome measurement process, each
physical qubit in the memory can suffer an error with an independent
probability $p$ and each measurement outcome can be incorrect with
probability $q$.  For the case of toric codes \cite{kitaev:03},
which were the first example of topological codes, error correction
($q=0$) can be studied via the two-dimensional (2D) random-bond Ising
model \cite{edwards:75,nishimori:01}.  Including faulty measurements
($q>0$), the error process is mapped onto a 3D random-plaquette gauge
model \cite{dennis:02-ea,ohno:04-ea}.

\begin{figure}
\psfrag{a}{\large a}
\psfrag{b}{\large b}
\psfrag{c}{\large c}
\psfrag{d}{\large d}
\psfrag{e}{\large e}
\psfrag{f}{\large f}
\psfrag{g}{\large g}
\psfrag{H}{\large $\mathcal H$}
\psfrag{T}{\large $\mathcal T$}
\center
\includegraphics[width=.9\columnwidth]{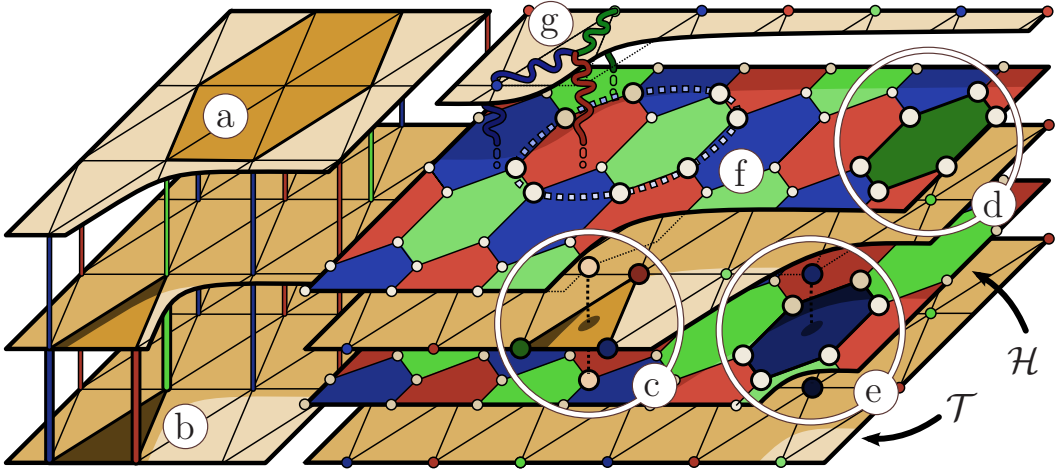}
\vspace*{-0.3cm}
\caption{
Left: Lattice representing the error history. Right: Derived
lattice for the lattice gauge theory model. Time corresponds to
the vertical axis.  The vertices of the $\mathcal{T}$ layers and
the hexagons of the $\mathcal{H}$ layers are three-colored.  (a,b)
Elementary equivalences between error histories.  (c,d) Hamiltonian
terms (five-body and planar hexagons). (e) Flipping the encircled
spins is a gauge symmetry. (f) Horizontal Wilson loop. In this work
we study elemental loops encircling one hexagonal plaquette. (g)
Excitations are colored fluxes.}
\label{fig:lattice}
\end{figure}

More recently \cite{bombin:06,bombin:07} other approaches to
topological quantum error correction have been introduced. For
example, topological color codes (TCC) are a class of topological
codes that allow the transversal implementation of the Clifford
group of quantum gates \cite{bombin:06,bombin:07}. Remarkably,
these enhanced computational capabilities for quantum distillation,
teleportation, dense coding, etc.~are possible while preserving
a high error threshold \cite{katzgraber:09c-ea} (in comparison
to the Kitaev model \cite{kitaev:03}).  This result has been
obtained numerically after mapping the error correction ($q=0$) for
TCCs onto a random three-body Ising model on a triangular lattice
\cite{katzgraber:09c-ea}; later confirmed by other numerical methods
\cite{ohzeki:09a,wang:09-ea,landahl:09}.  It is, however, unclear
if quantum computations on TCCs can be performed reliably in the
presence of faulty measurements ($q\neq 0$).  Here we address this
problem by simulating a complex disordered lattice gauge theory (LGT)
\cite{comment:lgtbooks} -- a task that required the introduction of
novel tools to probe for an ordered phase. Our main results are:

$\Box$ A complete study of error correction in TCCs.  We find that,
for equal error qubit flip and measurement error rates $p=q$, the
error threshold is $p_c=4.8(2)\%$.

$\Box$ A novel Abelian lattice gauge theory with gauge group ${\mathbb
Z}_2 \times {\mathbb Z}_2$ and a peculiar lattice and gauge structure
that departs from the standard formulations of Wegner \cite{wegner:71}
and Wilson \cite{wilson:74}, see Fig.~\ref{fig:lattice}. We refer
to it as a {\em tricolored LGT}. Its structure reflects the error
history in color codes, rather than the discretization of a continuous
gauge theory.

$\Box$ A novel approach to pinpoint first-order phase transitions
in LGTs with disorder using the skewness of the average over Wilson
loop operators.

\section{Error Correction with Faulty Measurements}
\label{sec:qec}

To construct a topological color code, we consider a hexagonal
arrangement of the physical qubits on a 2D surface, such that the
dual lattice has a regular triangular form with 3-colorable vertices
(i.e., the vertices can be labeled with three colors such that adjacent
vertices have different colors). Note that each triangle $\triangle$ of
the dual lattice $\mathcal T$ (see Fig.~\ref{fig:lattice}) corresponds
to a physical qubit in the initial arrangement. We now attach to each
vertex $v$ two Pauli operators,
\begin{equation}
	X_v:=\otimes_{\triangle\ni v} X_{\triangle},\quad
	Z_v:=\otimes_{\triangle\ni v} Z_{\triangle}\,,
\end{equation} 
which act on the physical qubits associated with the six triangles
adjacent to $v$. These are called check operators because encoded
states $| \psi\rangle$ satisfy $X_v | \psi\rangle = Z_v |\psi\rangle =
|\psi\rangle$. Such states exist because the group generated by check
operators, called stabilizer group, does not contain $-\mathbf{1}$,
so that in particular check operators commute with each other.
The dimension of the encoded subspace depends only on the topology
of the surface where the code lives, something that is true in part
due to the colorability properties of the lattice. For example, a
regular lattice with periodic boundary conditions has the topology
of a torus and encodes two logical qubits  \cite{bombin:06}. As in
any topological code, a main feature is that local operators cannot
distinguish different encoded states.

To keep track of errors, check operators are measured repeatedly
over time, allowing for the detection of local inconsistencies with
the code.  Error correction amounts to guessing the correct error
history $E$ among those that are compatible with the measurement
outcomes. Indeed, many error histories have an equivalent effect, and
thus the ideal strategy is to compute which equivalence class $\bar E$
happened with the highest probability $P(\bar E)$. Therefore, error
correction is highly successful when for typical errors there is a
class that dominates over the others. Color codes, being topological,
have an error threshold below which the success probability approaches
unity in the limit of large codes.

Because color codes are Calderbank-Shor-Steane (CSS) codes
\cite{calderbank:96,steane:96a}, $X$ (bit-flip) and $Z$ (phase-flip)
errors can be corrected separately.  In doing so one ignores any
correlations between bit-flip and phase-flip errors, something that
can only make error correction worse.  Then, for simplicity, we assume
an error model where bit-flip and phase-flip errors are uncorrelated.
Before each round of measurements each qubit is subject to the channel

\begin{equation}
\rho \longrightarrow (1-p)^2 \rho + p (1-p) (X\rho X+Z\rho Z)+p^2 Y\rho Y,
\end{equation}
where $\rho$ represents the state of a single qubit and $p$ is both
the probability for a bit-flip error to occur and the probability for
a phase-flip error to occur.  As for the measurement outcomes of check
operators, each of them is wrong with independent probability $q$.

We focus on bit-flip errors, which change the eigenvalue of
all adjacent $Z_v$ check operators; $Z$ errors are analogous.
Identifying time with the vertical direction, we form a lattice of
stacked triangle layers, one per measurement round, with vertical
connections (see Fig.~\ref{fig:lattice}, left). Error histories are
represented as collections of triangles representing bit-flip errors
and vertical links that represent measurement errors.  If an error
history $E$ is composed by a total of $b$ bit-flip errors and $m$
wrong measurement outcomes, its probability is
\begin{equation}\label{prob}
	P(E)=(1-p)^{N-b}(1-q)^{M-m}p^bq^m
	\propto \left (\frac p{1-p}\right)^b \left (\frac q{1-q}\right)^m
\end{equation}
where $N$ is the total number of qubits and $M$ the total number of
check operators.

Following Ref.~\cite{dennis:02-ea}, we classify error histories $E$
in equivalence classes $\bar E$. Whenever two error histories $E$
and $E'$ belong to the same class and an error $E$ has happened,
the error correction is still successful even if we wrongly correct
according to $E'$.  The equivalence relation can be constructed from
two different kinds of elementary equivalences.  First, if $S$ is the
set of triangles meeting at a given vertex $v$ [Fig.~\ref{fig:lattice}
(a)] and the histories $E$ and $E\prima$ differ exactly on the elements
of $S$, then they are equivalent because the local operator $X_v$
has no effect on encoded states. In particular, if $P$ is a Pauli
operator representing an error history---including $X$ and $Z$
errors---and $\rho$ an encoded state, then
\begin{equation}
	X_v P\rho P^\dagger X_v = P X_v \rho X_v P^\dagger = P\rho P^\dagger. 
\end{equation}
We say that these elementary equivalences are of type I.
Second, suppose that $S$ is now the set including two triangles,
one on top of the other, and the vertical links connecting them
[Fig.~\ref{fig:lattice} (b)]. Then if the histories $E$ and $E\prima$
differ exactly on the elements of $S$, they are equivalent because two
unnoticed subsequent bit-flips on the same qubit are irrelevant. We
say that these elementary equivalences are of type II.

\section{Tricolored lattice gauge theory with randomness} 
\label{sec:rcgm}

As in Ref.~\cite{dennis:02-ea}, the first goal is to express the
probabilities of error classes in terms of the partition function of
a suitable statistical model. Our goal is to obtain
\begin{equation}
	P(\bar E):=\sum_{E\in \bar E} P(E) \propto
	Z_{E} (T):=\sum_{\sigma} e^{-\beta H_{E}(\sigma)}
\end{equation}
for a suitably parametrized Hamiltonian $H_E$.  We start by attaching
classical spins $\sigma=\pm 1$ to the elementary equivalences between
error histories discussed above, both of type I and II.  This produces
a new lattice [see Fig.~\ref{fig:lattice} (right)] with honeycomb
layers $\mathcal H$ sandwiched between triangular layers $\mathcal
T$. Spins on the $\mathcal T$-layers [$\mathcal H$-layers] correspond
to equivalences of type I [type II] while negative interaction
constants correspond to errors.  There are six-body interactions which
stem from measurement errors (vertical links in the previous lattice)
that involve the vertices of a given hexagon at a $\mathcal H$ layer,
see Fig.~\ref{fig:lattice}(d). There are also five-body interactions
due to bit-flip errors and thus related to triangles in ${\cal
T}$-layers: they involve the three vertices of the triangle plus
the two $\mathcal H$-vertices directly above and below the triangle,
see Fig.~\ref{fig:lattice}(c). We use the symbol $\tetra$ to denote
such sets of five vertices. The Hamiltonian of interest is
\begin{equation}
H_\gamma (\sigma) = 
-J \sum_{\tetra} \gamma_{\tetra} \prod_{j\in \tetra} \sigma_j 
- K \sum_{\hexagon} \gamma_{\hexagon} \prod_{k\in\hexagon}\sigma_k \, ,
\label{hamiltonian}
\end{equation}
where $J > 0$ and $K > 0$, $\gamma_{\tetra}, \gamma_{\hexagon}
= \pm 1$ and $j$ [$k$] runs over the vertices of each $\tetra$
[$\hexagon$].  Given an error history $E$, let $\gamma^E$ be such
that $\gamma^E_{\tetra}=-1$  when the triangle in $\tetra$ belongs to
$E$, and similarly for hexagons $\hexagon$ and their dual vertical
links. Also, let $\sigma_0$ represent the state with all spins in
the +1 state. Then
\begin{equation}\label{p1}
P(E)\propto e^{-\beta H_{\gamma^E}(\sigma_0)}
\end{equation}
if the
inverse temperature $\beta = 1/T$ as well as $J$ and $K$ satisfy
the conditions
\begin{equation}\label{Nishimori}
e^{-2\beta J}= p/(1-p)
\qquad \text{and} \qquad 
e^{-2\beta K}= q/(1-q)\, .
\end{equation}
Moreover, if two spin configurations $\sigma_1$ and $\sigma_2$
only differ by the sign of a single spin, and two error histories
$E_1$ and $E_2$ are equivalent up to the corresponding elementary
equivalence, then
\begin{equation}\label{p2}
H_{\gamma^{E_1}}(\sigma_1)=H_{\gamma^{E_2}}(\sigma_2) .
\end{equation}
Putting together Eqs.~(\ref{p1}) and (\ref{p2}) we obtain the desired
result
\begin{equation}
P(\bar E)\propto
Z_{\gamma^E} (\beta J):=\sum_{\sigma} e^{-\beta H_{\gamma^E}(\sigma)} .
\end{equation}

Following Ref.~\cite{dennis:02-ea}, the next step is to consider
a model with a Hamiltonian [Eq.~\eref{hamiltonian}] where the
parameters $\gamma$ are quenched variables, with $p$ and $q$
dictating the probability distribution of the $\gamma$'s such
that $P(\gamma^E):=P(E)$. That is, each $\tetra$ [$\hexagon$] has
negative sign with probability $p$ [$q$]. The resulting system has
four parameters, $\beta J$, $\beta K$, $p$ and $q$.  The connection
between error correction in color codes and this statistical model
happens along the sheet described by the Nishimori conditions,
Eqs.~\eref{Nishimori}. In that sheet, order in the statistical model
corresponds to errors being correctable \cite{dennis:02-ea}. For the
sake of simplicity, we set $p = q$, assuming the same fault rate for
qubit and measurement errors. In that case it is enough to consider
a statistical model with $K = J$, so that it has only two parameters,
namely $\beta J$ and $p$.  The critical $p$ along the Nishimori line
\begin{equation}
e^{-2\beta J}= p/(1-p)
\end{equation}
then gives the fault-tolerance threshold $p_c$ for color codes.

\section{Gauge Symmetry}
\label{sec:gauge}

The Hamiltonian in Eq.~(\ref{hamiltonian}) has a local symmetry per
hexagon $\hexagon$: flipping its spins and those above and below its
center [Fig.~\ref{fig:lattice}(e)] leaves the energy invariant. This
is a ${\mathbb Z}_2\times {\mathbb Z}_2$ gauge symmetry, as the
construction of suitable Wilson loops demonstrates.  Let us label
$\mathcal H$-hexagons according to the coloring of the $\mathcal
T$-vertices. Wilson loops come also in three colors.  For example,
a blue horizontal loop is the product of green and blue hexagonal
terms [Fig.~\ref{fig:lattice}(f)]. Since there are two independent
Wilson loops for a given surface with a single boundary, there are
four possible flux values going through it. Moreover, the total flux
through two regions is trivial if both have the same flux, showing
that the gauge is ${\mathbb Z}_2\times {\mathbb Z}_2$. Indeed,
excitations can be regarded as colored fluxes. First, depict an
excited hexagon as a vertical flux of the corresponding color. Second,
depict an excited triangle as the three merging fluxes (one of each
color). Using this convention, excitations take the form of closed
colored strings with branching points where three different colors meet
[Fig.~\ref{fig:lattice}(g)].

\section{Simulation Details} 
\label{sec:nsm}

The study of the partition function [Eq.~\eref{hamiltonian}]
constitutes a considerable numerical challenge. To compute the error
threshold including bit-flip and measurement errors, we need to compute
the $p\,$--$q\,$--$T_c$ phase diagram of the model. The special case of
equal error rates ($p=q$) studied here provides a useful guidance,
however, 23.4 CPU years ($6.4 \cdot 10^{15}$ operations on Brutus)
were needed to obtain the results in Fig.~\ref{fig:phase_diagram}.

\begin{table}[!tb]
\caption{
Simulation parameters: $L$ is the layer size, $M$ is the number
of layers, $N_{\rm sa}$ is the number of disorder samples, $t_{\rm
eq} = 2^{b}$ is the number of equilibration sweeps, $T_{\rm min}$
[$T_{\rm max}$] is the lowest [highest] temperature, and $N_{\rm T}$
the number of temperatures used.  \label{tab:simparams}}

\vspace*{1.0em}
\hspace*{6.5em}\begin{tabular*}{0.8\columnwidth}{@{\extracolsep{\fill}} l r r r r r r r}
\hline
\hline
$p$ & $L$ & $M$ & $N_{\rm sa}$ & $b$ & $T_{\rm min}$ & $T_{\rm max}$ &$N_{\rm
T}$ \\
\hline
$0.00$ & $6$, $9$ & $6$, $8$ & $1600$  &  $15$ & $1.20$ & $2.00$ & $64$\\
$0.00$ & $12$     & $12$     & $800$   &  $15$ & $1.20$ & $2.00$ & $64$\\
$0.00$ & $15$     & $14$     & $400$   &  $17$ & $1.20$ & $2.00$ & $64$\\
$0.02$ & $6$, $9$ & $6$, $8$ & $1600$  &  $16$ & $0.90$ & $1.80$ & $52$\\
$0.02$ & $12$     & $12$     & $800$   &  $17$ & $0.90$ & $1.80$ & $52$\\
$0.02$ & $15$     & $14$     & $400$   &  $19$ & $0.90$ & $1.80$ & $52$\\
$0.03$--$0.039$ & $6$, $9$ & $6$, $8$& $1600$ & $17$ & $0.70$ & $1.40$ & $52$\\
$0.03$--$0.039$ & $12$     & $12$    & $800$  & $19$ & $0.70$ & $1.40$ & $52$\\
$0.03$--$0.039$ & $15$     & $14$    & $400$  & $21$ & $0.70$ & $1.40$ & $52$\\
$0.04$--$0.060$ & $6$, $9$ & $6$, $8$& $1600$ & $18$ & $0.50$ & $1.20$ & $52$\\
$0.04$--$0.060$ & $12$     & $12$    & $800$  & $20$ & $0.50$ & $1.20$ & $52$\\
$0.04$--$0.060$ & $15$     & $14$    & $400$  & $22$ & $0.50$ & $1.20$ & $52$\\
\hline
\hline
\end{tabular*}

\end{table}

Because the model is a LGT the natural (gauge invariant) order
parameter is the Wilson loop.  Due to the presence of disorder
and the complexity of the lattice, a clean scaling analysis using
the area/perimeter laws \cite{ohno:04-ea} is imprecise because
large Wilson loops show strong corrections to scaling. Instead,
we investigate in detail the smallest horizontal loops in
the system, 
\begin{equation}
W_{\hexagon}:=\prod_{k\in\hexagon}\sigma_k ,
\end{equation}
which correspond to single hexagon plaquettes $\hexagon$ and
record their average value over all loops in the system, i.e.,
\begin{equation}
w =\frac{1}{N_{\hexagon}}\sum_{\hexagon}W_{\hexagon} , 
\end{equation}
suitably averaged over Monte Carlo time and disorder. Without
disorder ($p=0$) this order parameter shows a clear signal of a
transition. However the detection of the transition temperature
$T_c$ becomes increasingly difficult when a finite error probability
($p>0$) is introduced.  Therefore, we also study the whole distribution
$f(w)$. For a first-order phase transition we expect a characteristic
double-peak structure to emerge near the transition.  To better
pinpoint the transition, we introduce a derived order parameter $\zeta$
related to the skewness of the distribution
\begin{equation}
\label{eq:skewness}
\zeta(\tilde w) = \av{\aT{\tilde w^3}}/\av{\aT{\tilde w^2}}^{3/2}\, ,
\end{equation}
where $\tilde w = w-\av{\aT{w}}$, $\aT{\cdots}$ denotes a
thermal average and $\av{\cdots}$ represents an average over
the disorder. The skewness of a symmetric function is zero and
becomes positive (negative) when the function is slanted to the left
(right). Therefore we can identify the point where $\zeta$ crosses
the horizontal axis with an equally-weighted double peak structure in
$f(w)$, i.e., $\zeta(T = T_c) = 0$.  We have compared these results
to the peaks in the specific heat as well as a Maxwell construction
and find perfect agreement.  To compute the $p\,$--$T_c$ phase diagram,
we fix the value of the error rate $p$ and vary the temperature $T$
until we detect a Higgs-to-confining transition for $T = T_c$.
$T_c(p)$ is then the critical line separating both phases. The error
threshold $p_c$ is given by the crossing point between $T_c(p)$
and the Nishimori line \cite{nishimori:81}.

The term ``Higgs-to-confining'' refers to the transition from one
gapped phase to another in a LGT. These gapped phases are characterized
by non-vanishing closed string correlators, the Wilson loops, and
can be distinguished by investigating the perimeter and area laws,
respectively.  The Higgs phase corresponds to a perimeter law for
the Wilson loop correlator, regardless of whether an explicit Higgs
field is present in the theory. The rationale is that static matter
sources, when introduced in the Higgs phase, will become unbounded,
i.e., deconfined, as opposed to the confinement which occurs in the
phase with an area law. Therefore, here the Higgs phase refers to the
(magnetically) ordered phase, while the confined phase corresponds
to the disordered phase found for larger values of $p$ and $T$.

It has been previously shown \cite{dennis:02-ea}, that the error
threshold is given by the crossing point of the Nishimori line with
this phase boundary. We recall that the Nishimori line $T=T(p)$ is
the locus describing a quantum computer in the presence of external
noise, while the rest of the phase diagram is merely introduced as
an auxiliary tool to locate the multicritical point.  Note that on
the Nishimori line, the effects of thermal fluctuations and quenched
randomness are in balance. For weak disorder $p$ and low temperatures
$T$, the system is in a magnetically ordered Higgs phase. In terms
of the color code, this indicates that all likely error histories
for a given error syndrome are topologically equivalent and error
recovery is achievable.  However, at a critical disorder value $p_c$
(and a corresponding temperature determined by Nishimori's relation),
magnetic flux tubes condense and the system enters the magnetically
disordered confinement phase. In this case, magnetic disorder means
that the error syndrome cannot point to likely error patterns belonging
to a unique topological class; therefore the topologically encoded
information is vulnerable to damage.

For the simulations we study systems of size $L^2\times M$ with $M$
along the vertical direction. Periodic boundary conditions are applied
to reduce finite-size effects. The colorability and stacking of the
layers requires that $L$ [$M$] is a multiple of $3$ [$2$]. For the
system to be as isotropic as possible, we choose parameter pairs such
that $M\in\{L,L-1\}$. Because the numerical complexity of the system
increases considerably when $p > 0$, we use the parallel tempering
Monte Carlo technique \cite{hukushima:96}.  Simulation parameters
are listed in Table \ref{tab:simparams}. Equilibration is tested by
a logarithmic binning of the data. Once the last three bins agree
within statistical error bars, the system is deemed to be in thermal
equilibrium.

\begin{figure}
\center
\includegraphics[width=0.7\columnwidth]{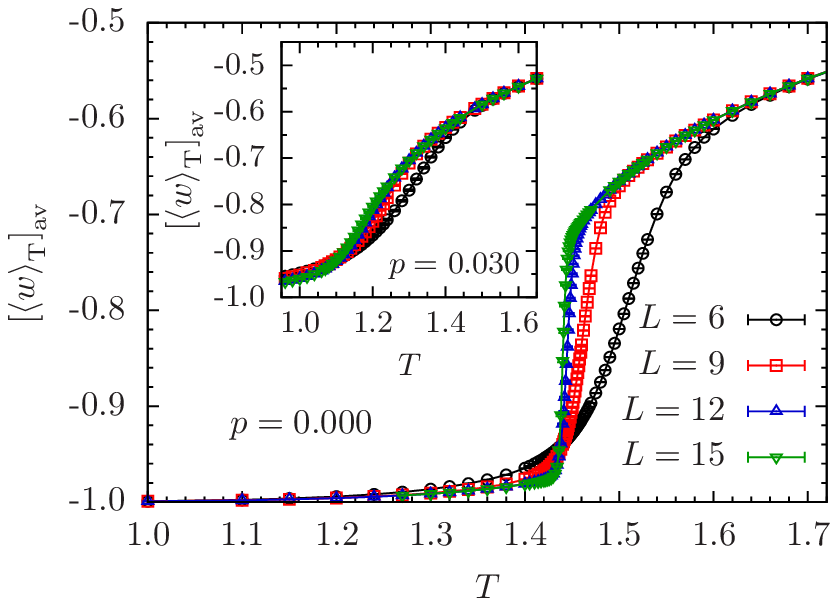}
\vspace*{-0.3cm}
\caption{ 
Average Wilson loops for $p = 0$ as a function of temperature for
different system sizes $L$. For increasing $L$ a sharp drop develops
at the transition. The inset shows data for $p = 0.03$.
}
\label{fig:wilson_average}
\end{figure}

\begin{figure}
\center
\includegraphics[width=0.7\columnwidth]{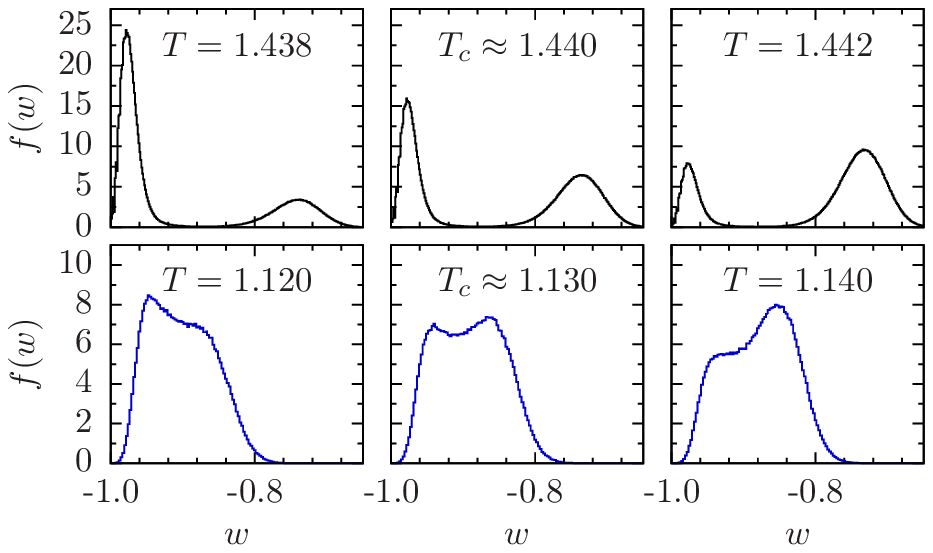}
\vspace*{-0.3cm}
\caption{
Wilson loop histograms $f(w)$ for systems of size $L=15$
with $p=0$ (top row) and $p=0.03$ (bottom row, blue). The panels show
data for temperatures slightly below (left), at (center), and above
the transition (right). The double-peak structure is indicative
of a first-order transition. For $T \approx T_c$ the peaks have
equal weight, whereas for $T < T_c$ ($T > T_c$) it is slanted to
the left (right), thus motivating the use of the skewness $\zeta$
(Fig.~\ref{fig:wilson_skewness}) as an order parameter. All panels
have the same horizontal scale.}
\label{fig:wilson_histograms}
\end{figure}

\section{Results} 
\label{sec:results}

\begin{figure}
\center
\includegraphics[width=0.7\columnwidth]{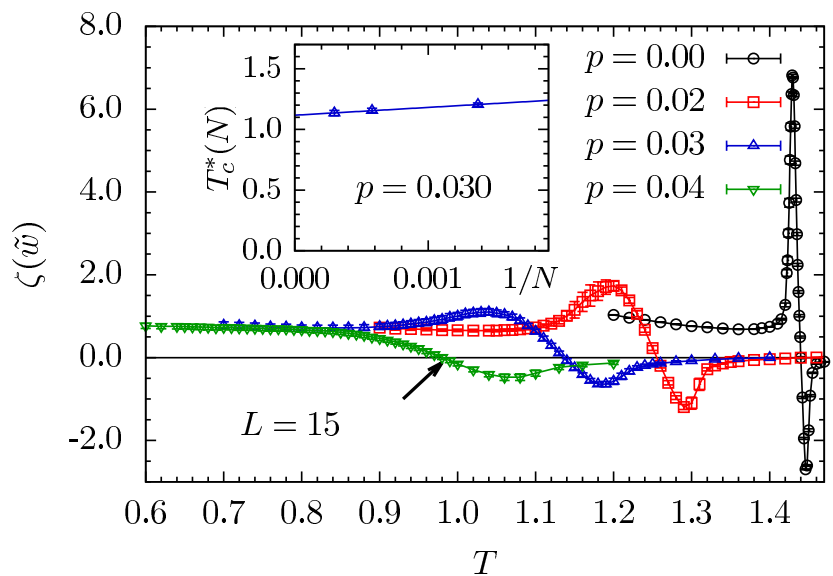}
\vspace*{-0.3cm}
\caption{
Skewness $\zeta$ for different disorder rates $p$ and $L=15$
as a function of temperature $T$. The crossing points with the
horizontal axis (arrow) correspond to where both peaks in the Wilson
loop histograms have equal weight, thus signaling the transition.
Inset: Example extrapolation of $\zeta[\tilde w,T = T_c^*(N)] = 0$
to the thermodynamic limit for $p=0.03$.}
\label{fig:wilson_skewness}
\end{figure}

\begin{figure}
\center
\includegraphics[width=0.7\columnwidth]{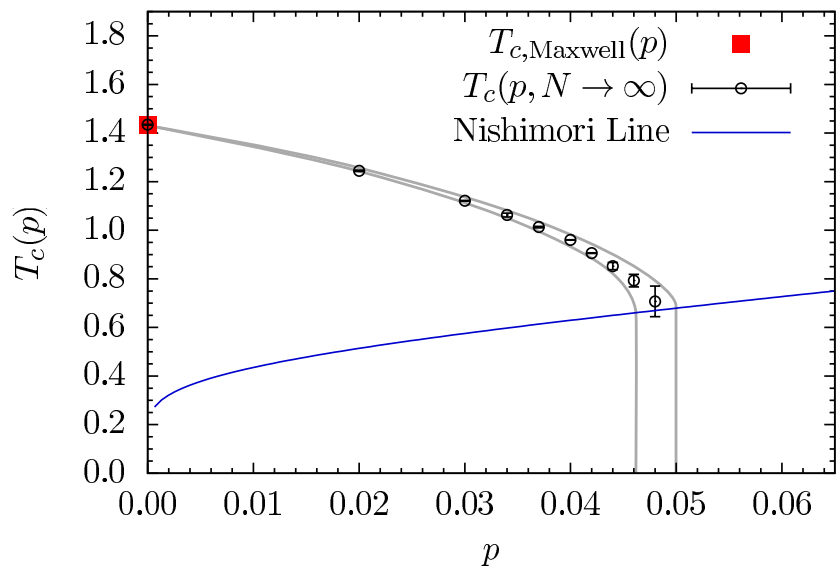}
\vspace*{-0.3cm}
\caption{
$p\,$--$T_c$ phase diagram. For $p > p_c = 0.048(2)$ errors cannot
be recovered.  The thin (blue) line represents the Nishimori line.
In the regime marked by dashed lines the determination of $T_c(p)$
is difficult. The red square is computed using a Maxwell construction
of the Wilson loop order parameter and agrees perfectly with the
estimate from the skewness.}
\label{fig:phase_diagram}
\end{figure}

Figure \ref{fig:wilson_average} shows the average Wilson loop value
as a function of temperature for the pure system ($p = 0$) and for
an error rate of $p=0.03$ (inset). For the system without disorder,
the transition is clearly visible (note that extrapolating to the
thermodynamic limit shows that for $p = 0$ the transition seems to also
be first order. See also Fig.~\ref{fig:wilson_skewness}). However, when
$p > 0$ it is difficult to determine the location of the transition. An
alternative view is provided by the histogram of Wilson loop values
$f(w)$ for different temperatures (Fig.~\ref{fig:wilson_histograms}).
Below $T_c$ (left panel), one can observe the development of a
shoulder that eventually becomes a second peak. The two peaks have a
symmetric weight distribution at the transition temperature (center
panel). Above $T_c$ the initial peak starts to disappear (right panel)
and the distribution slants to the right.  This property is mirrored
by the skewness of the distribution (Fig.~\ref{fig:wilson_skewness}).
$\zeta(\tilde w)$ has a positive [negative] peak where the second
[first] peak in $f(w)$ develops [disappears], with a zero where the
weight distribution is symmetric, i.e., at $T = T_c$. We have compared
our results using the skewness to the peak position of the specific
heat, as well as a Maxwell construction. Not only does the skewness
deliver the most precise results, for large disorder ($p \gtrsim
0.03$) it is the {\em only} method that reliably shows a sign of a
transition, if present. We estimate $T_c$ in the thermodynamic limit
by plotting the size-dependent crossing point $T_c^*(N)$ against $1/N$
and applying a linear fit (inset). The full phase diagram is shown
in Fig.~\ref{fig:phase_diagram}; the crossing between  $T_c(p)$ and
the Nishimori line (thin blue line) determines the error threshold
$p_c=0.048(2)$. Finally, we would like to note that both vertical and
horizontal Wilson loops give $T_c$ values that agree within error bars.

\section{Conclusions} 
\label{sec:concl}

By mapping topological color codes with both qubit and measurement
errors onto a classical statistical mechanical model with disorder, we
have obtained an Abelian lattice gauge theory with a peculiar lattice
and gauge structure. To date, toric codes were the only codes whose
complete fault-tolerance properties had been studied. Using Monte Carlo
simulations we estimate the error threshold for topological color codes
with both bit-flip and measurement errors to be $p_c=4.8(2)\%$ (to be
compared with $p_c \approx 3\%$ for the toric code \cite{ohno:04-ea}).
To obtain this result we introduce a new approach that uses the
skewness of the Wilson loop distribution to pinpoint first-order
phase transitions in lattice gauge theories with disorder.

\ack

We thank A.~P.~Young for useful discussions.  M.A.M.-D.~and
H.B.~acknowledge financial support from research grant QUITEMAD
S2009-ESP-1594, FIS2009-10061, UCM-BS/910758 and EU grant PICC.
H.G.K.~acknowledges support from the SNF (Grant No.~PP002-114713).
The authors acknowledge ETH Zurich for CPU time on the Brutus cluster
and the Centro de Supercomputaci\'on y Visualizaci\'on de Madrid
(CeSViMa) for access to the Magerit cluster.

\bibliographystyle{unsrt}
\Bibliography{99}
\nonum
\endbib
\bibliography{refs,comments}

\end{document}